# The Quest and Hope of Majorana Zero Modes in Topological Superconductor for Fault-Tolerant Quantum Computing: An Introductory Overview

Nur R. Ayukaryana[1, a)], Mohammad H. Fauzi[2], Eddwi H. Hasdeo[2,3, b)]

[1]*Engineering Physics Department, Faculty of Industrial Technology, Institut Teknologi Bandung, Indonesia*
[2]*Research Center for Physics, Indonesian Institute of Sciences, South Tangerang, Indonesia*
[3]*Department of Physics and Material Science, University of Luxembourg, Luxembourg*

[a)]Corresponding author: ayukaryana@students.itb.ac.id
[b)]eddw001@lipi.go.id

**Abstract.** Ettore Majorana, in his short life, unintendedly has uncovered the most profound problem in quantum computation by his discovery of Majorana fermion, a particle which is its own anti-particle. Owing to its non-Abelian exchange statistics, Majorana fermions may act as a qubit for a universal quantum computer which is fault-tolerant. The existence of such particle is predicted in mid-gap states (zero modes) of a topological superconductor as bound states that have a highly entangled degenerate ground state. This introductory overview will focus on the simplest theoretical proposals of Majorana fermions for topological quantum computing in superconducting systems, emphasizing the quest from the scalability problem of quantum computer to its possible solution with topological quantum computer employing non-Abelian anyons on various platforms of certain Majorana fermion 'signature' encountered.

## INTRODUCTION

The classical computer which employs bit is the technology we encounter nowadays. Their information is encoded in binary numbers which have value of 0 or 1 and the physical object carrying the information is the "bit", it is represented by the state of a transistor inside a silicon chip. At microscopic level of atoms and electrons, the world obeys the law of quantum mechanics. One of its fundamental concepts is superposition. With superposition, it is possible to create a system that has two basic states known as "qubit". Qubit is a quantum analogue for a "bit". One qubit can be in a superposition of two basic states of 0 and 1,

$$|\psi\rangle = a|0\rangle + b|1\rangle. \tag{1}$$

To specify the state $|\psi\rangle$ of the qubit, one needs to define $a$ and $b$. For instance, a system with two qubits will have a superposition of four basic states,

$$|\psi\rangle = a|00\rangle + b|01\rangle + c|10\rangle + d|11\rangle. \tag{2}$$

Adding $N$ qubit into the system results in $2^N$ possible basic states. Hence, the more qubits one has, the more entangled quantum computer codes proliferate. One requires to specify what superposition of the state is, in order to perform quantum computation. However, not all information is accessible to be quantum measured. Therefore, one needs to develop quantum algorithm to explore the huge amount of quantum superposition but leave the system in one of the basic states which can be detected with certainty.

Besides arbitrary superpositions of 0 and 1 within each qubit, quantum computer is able to support entanglement. In closer considerations, Josza [1] showed that an entangled state is a superposition of product states which cannot be expressed as a single product state. For example, one can create an entangled state of two spins with opposite direction. If one separates these two spins after creating the entangled state, then the spin properties will not change no matter how far the measurement takes place. However, these entangled states are very fragile and can be easily destroyed by unwanted interference. Hence, building the true universal quantum computer requires so much efforts and investments.

The development of quantum computing perhaps began from Feynman in 1982 [2] when he stated that the simulation of quantum evolution in classical computers facing unavoidable exponential time slowdown. He implicitly asked whether one may use quantum mechanics in a computer to compute more efficiently than on a classical computer. Although, he did not specify the context to the extent of quantum computer hardware. In 1985, Deutsch [3] was the first to ask the question explicitly. He proved that if 2-state systems could be made to evolve, then any unitary evolution could be produced, and therefore the evolution could be made to simulate any physical systems [4]. This simple operation is now called quantum gates because they are analogous of logic gates in classical computers. Quantum gates manipulate the qubits to perform a computation.

Generally, quantum computing involves three major steps: initialization, unitary evolution, and measurement. To perform quantum computation, one needs to initialize the state of qubits at the beginning as input to the calculation. Then, perform arbitrary controlled unitary operations on this initial state. Finally, measure the qubits as the output. A universal quantum computer utilizes a sufficiently large set of gates to perform arbitrary quantum algorithms. His proposal is then widely considered as a blueprint for quantum computing. Since then, a major development and breakthrough has been achieved in the field, though there are still several problems that have not been solved.

## Future of Computing and Electronics

Quantum computer has been a real hype for the past few years. The boost is not only coming from academia, but also from big tech companies and even several leading countries such as the United States and China. IBM, Google, Intel and Microsoft invest largely for its development, not to mention any other start-ups e.g., D-Wave Systems, Rigetti Computing, and IonQ. The most attractive algorithms for applications of quantum computation are probably factorization algorithm developed by Peter Shor [5], Grover's search algorithm [6] and quantum simulation for computational chemistry. Shor's result showed that factoring and extraction of discrete algorithm are both solvable in quantum polynomial time. This is a significant threat for most popular public-key encryption system available nowadays including RSA (Rivest, Shamir, Adelman) algorithm, and others based on the integer factorization problem and the elliptic-curve discrete logarithm or discrete logarithm problem [7]. They can be solved fast by Shor's algorithm [5]. Although, such remarkable ideas of internet code breaking will not present in the 50 qubits quantum computer which is now available.

In addition, quantum computers can speed up exhaustive search as proposed by Lov Grover in 1996. Grover's algorithm provides a quadratically faster search than unoptimized classical computer algorithm for an unsorted database. It can be thought of as looking for a particular telephone number in the telephone directory for someone with unknown name. In classical algorithm, one requires to search through the list on average $N/2$ steps for a list of $N$ items. However, Grover's algorithm only needs of order $\sqrt{N}$ steps. The proposal is quite advance although it cannot outperform classical computer by an order of exponentially faster. Furthermore, Bennett *et al.* [8] later showed that the algorithm cannot resolved NP-hard problem that is classically hard and quantumly hard. It is important to emphasize that there is a problem which is classically hard but quantumly easy i.e., the task of simulating a many-particle quantum system.

Computational chemistry is a major interest in many fields including chemistry, material science, solid-state physics, condensed matter physics, biophysics, and biochemistry. However, it is limited by exponential increase of resources along with increasing problem size. Thus, nowadays, the simulation in digital computer is limited to finding the ground state energy of extremely small collections of molecules. In quantum computing scheme, this problem may be addressed to scale polynomially and provide speedup as well as enable simulations of larger molecules and excited energy system [9]. Colless *et al.* [10] demonstrated a complete energy spectrum of $H_2$ molecule with near chemical accuracy using VQE (Variational Quantum Eigensolver). Kandala *et al.* [11] showed that a VQE on six-qubit superconducting quantum processor may address molecular problems beyond period I elements, up to $BeH_2$. Hence, with the advancement of quantum computers for the next few decades, the future electronics can be predicted precisely and thus reducing trial-and-errors in electronics fabrication.

Moreover, a recent hot development in quantum computing is quantum machine learning with parameterized quantum circuits. In the future, the real-world learning tasks with real advantage of this idea will be a key issue and interesting task in quantum computing. Despite of those algorithms, the fact that quantum computer cannot be constructed in classical computers are amusing. More accurately, people do not know how to simulate quantum computer using a digital computer. Recent implementation of quantum computing is through cloud-based system and companies have enabled their resources for public such as Qiskit by IBM.

## Challenges in Quantum Computers

Nowadays, the quantum computing technology is named as NISQ (*Noisy Intermediate Scale Quantum*) [12] which refers to intermediate scale size of quantum computer that will be available in the next few years with 50 to hundred qubits. Quantum computer can be implemented with various quantum systems such as trapped ions, superconducting qubits, photons, and silicon. With the recent quantum computer hardware for controlling trapped ions or superconducting qubits, the error rate per gate for two-qubit gates is above 0.1% level and often worse [12]. Furthermore, it is not known whether the error rate will be that low along with the increasing number of qubits.

The largest obstacle in quantum computer is error in the form of inaccuracy and decoherence. Analogue to classical computation, small errors can accumulate over time and eventually add up to large errors. Also, to correct an error, one should acquire some information about the error by making measurement. In classical computation, they are usually corrected by redundancies (keeping multiple copy of information and checking them). In quantum computer, quantum information cannot be copied [13] and the measurement of quantum state during calculation will collapse the wavefunction and destroy quantum superposition, in which measurement will result in gibberish.

To solve inaccuracy, Peter Shor [14] and Andrew Steane [15] developed the first quantum error correction (QEC) protocol. Yet, the QEC protocol itself is a complex task which requires encoding and recovery of quantum information. The QEC scheme is implemented when one attempts to process quantum information. The error in one qubit will propagate to other qubits while they are interacting with each other to perform quantum gates operation. In 1996, Knill and Laflamme [16] showed that there is an accuracy threshold for storage of quantum information. Then, in 2000, Gottesman [17] revealed that if the error is below a certain value, it is possible to do a long quantum computation with negligible error. Therefore, the quantum gates must be designed to minimize this error in certain tolerance.

As the first problem is solved by QEC protocol with reliable accuracy threshold on each quantum gate, the second problem, decoherence, has been solved theoretically by Shor's discovery in the concept of fault-tolerant quantum computation [18]. Furthermore, decoherence can be expressed purely in terms of inaccuracies in the state of quantum system interacting with auxiliary state which is the environment. Consequently, the decoherence-reduction method can be used to treat inaccuracies, and vice versa. According to Calderbank and Shor [19], the use of "good quantum error correcting codes" can reduce both decoherence and inaccuracy while performing computations in quantum data. This concept is then developed by Alexei Kitaev in 2003 using anyons [20]. In this scheme, the error is said to be corrected at physical or hardware level. Furthermore, the idea is refined with the idea of topological quantum computation favored by Majorana fermion as the qubit [21].

In other words, from practical standpoint, the errors can be divided into two kinds. First, error while qubit is processed or manipulated. Second, error while qubit is storing quantum information. In quantum information processing, for example, rotating qubit by 90.01° rather than 90° or unitary error is an issue of how one can manipulate the system. In contrast, while qubit is storing quantum information, the error arises due to the qubit interaction with environment.

## Topological Quantum Computer

The general concept of quantum error correction and topology lies in the similar nature, storing and manipulating qubit in "global" form. Thus, any local disturbances will not affect the overall system. However, in the NISQ era, the entangled states are not robust enough against local disturbances. Therefore, there are two major focuses for achieving quantum supremacy. First, to find new algorithms beyond discovered algorithms of searching and factorizing which will outperform classical computation. Second, to perform quantum computations which is resilient to errors, both inaccuracy and decoherence, and thus overcome scalability problem. In fact, the demonstration of quantum error correction with sufficiently isolated quantum systems and sufficiently precise quantum gates for reducing decoherence effects can allow fault-tolerant quantum computation. Nevertheless, the required thresholds are too stringent. Consequently, it is unclear whether qubit-based quantum computation can be made fault-tolerantly only by QEC scheme. Rasetti and Castagnoli [22] was the first to argue that anyon could be employed to perform quantum computation.

The idea of statistical mechanics in anyons was originated from Arovas *et al.* in 1985 [23] and was previously studied by Frank Wilczek in 1982 [24]. Wilczek mentioned that because the interchange of two particles orbiting around magnetic flux tubes can give *any* phase between bosons ($e^0$) and fermions ($e^{i\pi}$), he called it *anyons*. In three spatial dimensions, particles can be classified as bosons and fermions due to their fundamental property. Bosons can occupy the same states, but fermions can only occupy different states with each particle stacked together. In two spatial dimensions, the possibility of quantum statistics is not limited to boson and fermion, but rather allow continuous interpolation between these extremes. The quantum statistics is determined by the phase of the amplitude associated with slow motion of distance particles around one another. In statistical evolutions, the phase factor evolutions between fermionic case ($e^{i\theta} = -1$) and bosonic case ($e^{i\theta} = +1$) are known as Abelian anyons. Abelian anyons provide an example of unitary transformation that can be performed exactly in topological states of matter. Unfortunately, it is a trivial transformation, only changing the phase of wave function. To apply unitary transformations which will be useful for universal quantum computation, one requires a special class of topological states that support non-Abelian anyons. According to Kitaev [20], non-Abelian anyons undergo a non-trivial unitary transformation when one particle moves around another. This motion is termed as braiding and a measurement of its final state is done by joining the particles in pairs and observing the result of fusion.

This anyonic quantum computation scheme is then known as topological quantum computer. It encodes and manipulates information by employing anyons as qubits. The computational power of these anyons lies in their fusion and braiding property. The property is in the form of particle statistics that is resilient against environmental perturbations, thus implying that quantum information can be encoded fault-tolerantly. Fusion corresponds to bringing two anyons together and determines their behavior collectively. It can be viewed as putting two anyons in a box and identifying the statistical behavior of the box without paying attention to what happens inside the box. Generally, fusion rules are written as:

$$a \times b = N_{ab}^c c + N_{ab}^d d + \cdots \quad (3)$$

The rules indicate the possible outcomes of $c, d, ...$, listed with the $+$ symbol that result when anyons $a$ and $b$ are brought together which is denoted by the $\times$ symbol. The ordering of $a$ and $b$ is not important, thus

$$a \times b = b \times a. \quad (4)$$

When $a$ and $b$ are fused together, they produce particle $c$ by several distinct mechanism which then enumerated in $N_{ab}^c$. It is possible to manipulate fusion of two anyons in a particular way so that they have unique fusion outcome. The fusion outcome in equation (3) can be understood as different possible preparations of $a$ and $b$ that would result to a certain fusion outcome $c$.

Finally, anyons are systematically characterized by their fusion behavior. For instance, Abelian anyons only have a single fusion channel

$$a \times b = c. \quad (5)$$

Their fusion space is one-dimensional. In the other hand, non-Abelian anyons always have multiple fusion channels that give rise to higher dimensional fusion spaces

$$\sum_c N_{ab}^c > 1. \quad (6)$$

This simple property is closely related to their statistical behavior as previously described. Non-Abelian statistics is manifested as a non-trivial evolution between the different possible fusion outcomes given in (3). In contrast, Abelian statistics corresponds to the evolution of a unique state by a phase factor.

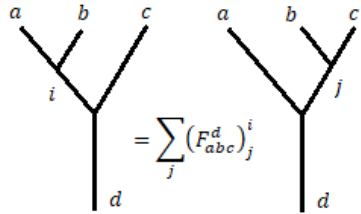

**FIGURE 1.** The order of fusion processes between three anyons.

When fusing several anyons, one may freely choose the order of basic fusion processes. This is illustrated in Fig. 1, three anyons $a, b,$ and $c$ with total fusion channel $d$ can be fused in two different ways which are fusing $a$ and $b$ with outcome $i$ and fusing $b$ and $c$ with outcome $j$. These two fusions result in the same fixed total fusion outcome, $d$. The matrix $F_{abc}^d$ with $i, j$ elements

$$\left(F_{abc}^d\right)_j^i \quad (7)$$

that relates those two different fusion processes is called the fusion of $F$ matrix. The dimensionality of this matrix depends on the number of possible in-between outcomes of the fusions. The choice of fusion order is a degree of freedom in the description of several anyons. Choosing the order in which anyons are fused can be viewed as a choice of basis and the $F$ matrix as a transformation between different bases [25].

The space of states which corresponds to the fusion process is the Hilbert space of anyons. Denoting the fusion Hilbert space of $n$ anyons by $M_{(n)}$, the fusion of Abelian anyons is trivial

$$dim(M_{abelian}) = 1. \tag{8}$$

since Abelian anyons have only a single fusion outcome. In contrast, the Hilbert space of two initial non-Abelian anyons $a$ and $b$ with a fusion outcome $c$ with multiplicity $N_{ab}^c$ gives rise to equally many states and thus the Hilbert space is given by

$$dim(M_{(3)}) = N_{ab}^c. \tag{9}$$

If one adds anyon $c$ such that the outcome is $d$, one can initially fuse $a$ and $b$ then fuse the outcome $i$ of this fusion with $c$ in order to get $d$, this fusion process may be described as

$$|i\rangle = |a, b \to i\rangle |i, c \to d\rangle = \sum_j (F_{abc}^d)_j^i |b, c \to j\rangle |a, j \to d\rangle. \tag{10}$$

Alternatively, one may consider fusing $b$ and $c$ and their outcome $j$ with $a$ to obtain $d$, changing between those two different fusion states correspond to the $F$ move described in Fig. 1. This equation then can be put simply

$$|i\rangle = \sum_j (F_{abc}^d)_i^j |j\rangle. \tag{11}$$

Considering more anyons to the system for ordering $n$ anyons $a_i$ with $i = 1, \ldots, n$, one has states

$$|e\rangle = |e_1, e_2, \ldots, e_{n-3}\rangle = |a_1, a_2 \to e_1\rangle|e_1, a_3 \to e_2\rangle \ldots |e_{n-3}, a_{n-1} \to a_n\rangle. \tag{12}$$

If one wishes to fuse the anyons in different order then one could employ the $F$ moves to transform the $|e\rangle$ states into the basis states of the new fusion order and therefore the number of different fusion probabilities is given by

$$dim(M_{(n)}) = \sum_{e_1 \ldots e_{n-3}} N_{a_1 a_2}^{e_1} \ldots N_{e_{n-3} a_{n-1}}^{a_n}. \tag{13}$$

Intuitively, the expression for $dim(M_{(n)})$ can be written in terms of the quantum dimension $d_i$ of anyon $i$. Quantum dimension is a fancy name that refers to the dimension of the Hilbert space associated to an anyon. Starting from the fusion rules of non-Abelian anyons $a \times b = \sum_c N_{ab}^c c$, one may define the quantum dimension

$$d_a d_b = \sum_c N_{ab}^c d_c. \tag{14}$$

Abelian anyons, such as vacuum, always have $d_i = 1$, while non-Abelian anyons necessarily have $d_i > 1$.

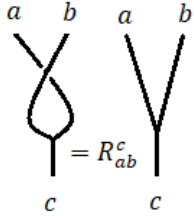

**FIGURE 2.** The clockwise exchange of anyons $a$ and $b$ with outcome $c$ gives phase of $R_{ab}^c$

Furthermore, the statistical evolutions of the effect of exchanging two anyons, $a$ and $b$, is shown in Fig. 2 when their fusion channel is fixed, i.e., $a \times b \to c$. This exchange can be viewed as a half twist of the particle $c$. Therefore, the exchange evolution $R_{ab}^c$ of the fusion state $|a, b \to c\rangle$ would simply be a phase factor as it corresponds to the rotation of a single particle. The exchange matrix $R_{ab}$ is referred as $R$ matrix.

Finally, this description implies the statistical exchange of anyons that give rise to topological quantum computation. The scheme for anyonic quantum computation consists of three steps which follow the circuit quantum computation model. Firstly, the initialization of quantum system in a well determined quantum state is done by creating and arranging anyons. Secondly, quantum gates operation is performed by braiding of anyons. Lastly, the final state measurement for output is obtained from detecting anyonic charge.

Starting with a set of anyons that are prepared in a well-defined fusion state. For example, using pairs of non-Abelian anyons $a$ and $\bar{a}$ from the vacuum. The fusion state of these anyons belongs to a Hilbert space that increases exponentially with the number, $n$, of anyonic pairs, $dim(M_{(n)}) \propto d_a^n$. This Hilbert space always contains a subspace in which quantum information can be encoded in the usual way using qubit. Its dimension increases exponentially as a function of $n$. Thus, non-Abelian anyons are an efficient medium for storing quantum information. Having identified the logical encoding space, next, one needs to consider the gates that evolve under operations. Logical gates can be performed by braiding the anyons and evolving their fusion state by the $R$ matrix. In combination with $F$ matrices, one can evolve the encoded information in a non-trivial way. Ideally, one requires any arbitrary algorithm out of braiding anyons for achieving universal quantum computer. It can be realized with the $F$ and $R$ matrices having a dense set of unitary matrices acting on the qubits. At the end of the computation, one must measure the processed information which is encoded in the final fusion state of the anyons. It can be obtained by fusing the anyons in a series and retrieving the fusion outcomes $e_i$. As the fusion state of the anyons can be a superposition of many different basis states $|e_1, e_2, ...\rangle$, the measurement of final fusion state provides a probability distribution. This step constitutes the final readout of the computations.

To sum up, the fusion space evolution induced by anyon braiding does not depend on the details of the paths spanned by the anyons, instead only on their topology. Hence, an experimentalist implementing topological quantum computation may ignore the spanning of these paths as long as their global characteristics are realized. When anyons are kept far apart, the information encoded in the fusion space is not accessible by local operations. Thus, the environmental errors, acting as local perturbations to the Hamiltonian cannot alter the fusion states. This characteristic of anyons favors fault-tolerant quantum computation.

Last but not least, the question regarding to the physical realization of anyons is half-answered in concrete anyonic models namely Ising anyons and Fibonacci anyons. The later model corresponds to a scheme for supporting truly universal quantum computation only by braiding the anyons. The interest in the former model, Ising anyons, is due to the possible physical realization with near future technology. However, this model must be supplemented with simple dynamical phase rotations to support universal quantum computation. In the Ising anyon model, the particle types are the vacuum, 1, the non-Abelian anyon, $\sigma$, and the fermion, $\psi$. The fusion rules are given by
$$\sigma \times \sigma = 1 + \psi, \quad \sigma \times \psi = \sigma, \quad \psi \times \psi = 1 \tag{15}$$
with 1 fusing trivially with the rest of the particles, i.e., $\sigma \times 1 = \sigma$ and $\psi \times 1 = \psi$. In contrast, in Fibonacci model, there are only two types of anyons, the vacuum, 1, and the non-Abelian anyon, $\tau$. Thus, the only non-trivial fusion rule is
$$\tau \times \tau = 1 + \tau \tag{16}$$
This model turns out to support universal quantum computation. Meanwhile, the Ising model requires a phase gate in order to support universal quantum computation.

An example of the simplest non-Abelian excitation is the zero energy Majorana bound state. These systems are called topological since they possess the topological properties of anyonic exchange. Such topological systems carry localized excitation i.e., quasiparticles, that have highly entangled degenerate ground states. These highly entangled degenerate ground states differentiate between topological qubit and other qubits. A fermionic state at $E = 0$ with an exponentially large degeneracy of $2^N$ is produced by Majorana zero modes in $2N$ isolated vortices. A unitary evolution in this manifold is called braiding and a projective measurement is called fusion i.e., the process of bringing vortices together such that the zero modes overlap and split to allow fermion parity to be measured. In superconducting condensate, pairs of quasiparticles are absorbed as Cooper pairs. Their measurement outcome is an element of fused vortices which leave behind an unpaired or paired quasiparticle. This outcome is specified by the fusion rules. If two pairs of Majorana zero modes $\gamma_1, \gamma_2$ and $\gamma_3, \gamma_4$ are each in a state of definite fermion parity, then the fusion of one vortex from each pair will produce equal-weight superposition of even and odd fermion parity. In a formal notation, the fusion rule is expressed as
$$\gamma_2 \times \gamma_3 = 1 + \psi \tag{17}$$
where $\psi$ indicates the presence of unpaired fermion and 1 refers to the vacuum (no unpaired fermions).

In topological systems, the information is not accessible if those anyons are kept apart from each other and hence it is protected. Therefore, anyons can manipulate quantum information with very accurate quantum gates while maintaining the quantum information hidden at all times. Hence, two questions left unanswered in order to truly achieve universal quantum computing. First, to find suitable system with appropriate topological properties for realizing non-Abelian anyons employing zero energy Majorana bound states or so-called Majorana zero modes. Second, to carry out braiding operations necessary to achieve required unitary transformations.

## THEORETICAL BACKGROUD

### Majorana Fermions

In 1928, Paul Dirac proposed his relativistic wave equation for spin ½ particles [26]:

$$(i\gamma^\mu \partial_\mu - m)\psi = 0 \tag{18}$$

Dirac's equation connects the four components of a field ψ. The γ matrices needs to obey the rules of Clifford algebra, which is

$$\gamma^\mu \gamma^\nu \equiv \gamma^\mu \gamma^\nu + \gamma^\nu \gamma^\mu = 2\eta^{\mu\nu} \tag{19}$$

Dirac found a suitable set of 4x4 γ matrices that contains real and imaginary numbers. Thus, ψ must be a complex field to satisfy the equation. Dirac and other physicists mostly agree that the description of charged particles require complex field. In the language of quantum field theory, if a given field ϕ creates particle $A$ and destroys its anti-particle $\tilde{A}$, the complex conjugate $\phi^*$ will create $\tilde{A}$ and destroys $A$. Because electrons and positrons are distinct, the associated field ψ and $\psi^*$ must be different. Therefore, it can be seen that Dirac's equation predicted the notion of anti-particles. However, particles that are their own anti-particles associated with the fields must obey $\phi = \phi^*$. Hence, the fields must be real.

The question whether such description necessarily involves complex number is answered by Ettore Majorana in 1937 [27]. To construct an equation of Dirac's type (suitable for spin ½ particles) but capable of governing a real field, one needs γ matrices with purely imaginary numbers that satisfy Clifford algebra. Majorana discovered such matrices written as tensor products of usual Pauli matrices σ,

$$\widetilde{\gamma^0} = \sigma_2 \otimes \sigma_1 = \begin{pmatrix} 0 & 0 & 0 & -i \\ 0 & 0 & -i & 0 \\ 0 & i & 0 & 0 \\ i & 0 & 0 & 0 \end{pmatrix} \tag{20}$$

$$\widetilde{\gamma^1} = i\sigma_1 \otimes 1 = \begin{pmatrix} 0 & 0 & i & 0 \\ 0 & 0 & 0 & i \\ i & 0 & 0 & 0 \\ 0 & i & 0 & 0 \end{pmatrix} \tag{21}$$

$$\widetilde{\gamma^2} = i\sigma_3 \otimes 1 = \begin{pmatrix} i & 0 & 0 & 0 \\ 0 & i & 0 & 0 \\ 0 & 0 & -i & 0 \\ 0 & 0 & 0 & -i \end{pmatrix} \tag{22}$$

$$\widetilde{\gamma^3} = i\sigma_2 \otimes \sigma_2 = \begin{pmatrix} 0 & 0 & 0 & -i \\ 0 & 0 & i & 0 \\ 0 & i & 0 & 0 \\ -i & 0 & 0 & 0 \end{pmatrix} \tag{23}$$

Thus, Majorana's equation is simply a modification of Dirac's equation:

$$(i\widetilde{\gamma^\mu} \partial_\mu - m)\widetilde{\psi} = 0 \tag{24}$$

the $\widetilde{\gamma^\mu}$ matrices are purely imaginary and consequently $i\widetilde{\gamma^\mu}$ are real and the equation finally governs a real field $\widetilde{\psi}$. An electrically charged particle is different from its anti-particle because it has opposite electric charge. Electric charge is a measurable and stable property of a particle. For electrically neutral particle, however, it is possible to be its own anti-particle. The neutron which has spin ½ is not its own anti-particle because several neutrons can coexist within an atomic nucleus, but anti-neutron rapidly annihilates. Electrons and protons are neither. Consequently, the question of particle which is its own anti-particle remains unanswered.

## Majorana Fermions in Condensed Matter System

Ettore Majorana speculated that neutrino might satisfy his equation as particle which is its own anti-particle or Majorana fermion. However, the experimental discoveries of neutrino in 1956 [28] showed that the neutrino and anti-neutrino has strict distinction property and thus disapprove Majorana's idea. Also, the neutrino-less double β decay experiment which may validate the existence of Majorana fermion in cosmos has not been achieved yet. In a neutrino-less double β decay, two neutrons would be decaying into two protons and two electrons without emitting any neutrinos. It is only possible when the neutrino is its own anti-particle. In contrast, Majorana fermion received huge attentions in condensed matter system. It was first envisioned by Kitaev in 2003 [20], he argued that Majorana fermion enables to achieve fault-tolerant quantum computing because of its exotic exchange statistics namely non-Abelian statistics.

In condensed matter system, superconductivity changes the description of electrons and holes (half-integer spin particles). In superconductors, the absolute distinction between electrons and holes are blurry because electrons form Cooper pairs. Owing to the bosonic-like nature of Cooper pairs, they can form dense "condensate". As a consequence, electron number is no longer conserved: two electrons in a Cooper pair can be added or removed from the condensate without substantially changing its properties. The superconductor screens electric and confines magnetic fields so that charge is no longer observable. Physically, an electron mode (normal state) can lower its energy by mixing with a hole mode (normal state) attached to a Cooper pair.

There are certain types of superconductor in which Majorana-type excitations are predicted to emerge. Some superconductors contain magnetic flux tubes known as vortices. These vortices may trap so-called zero modes, the spin ½ excitons (bound states of electrons and holes) with very low (formally, zero) energy. Those zero modes are equal combination of particles and holes, and thus they are termed "partiholes". Such partiholes are different from conventional excitons which is always bosons, they are created by operator in the form of

$$\gamma_j = c_j^\dagger + c_j \tag{25}$$

with 'particle states' are associated with the creation operator $c_j^\dagger$ and anti-particle (hole) states with the conjugate operator, $c_j$. The partiholes operator, $\gamma_j$, create localized spin ½ particles that are their own anti-particle. For instance, the partiholes corresponds to zero modes which is called Majorana zero modes.

In fermionic state, a pair of Majorana fermions may be combined into $f = \gamma_1 + i\gamma_2$. Majorana fermion in this sense is half of a normal fermion, meaning that a fermionic state $c$ is obtained as a superposition of two Majorana fermions, $\gamma_1$ and $\gamma_2$. Each of Majorana fermion is basically split into a real and imaginary part of a fermion. This fermionic state (superposition of two Majorana fermions) represents conventional fermion but remains non-trivial because it is spatially non-local and cannot be addressed individually. In other words, they are spatially separated (prevented from overlapping) such that the fermionic state is protected from most type of decoherence. It cannot be changed by local perturbations affecting only one of its Majorana constituents. Hence, the fermionic operator, $f$, encodes a highly non-local entanglement which is resilient to decoherence. One can empty or fill the non-local state with no energy cost and resulting in ground state degeneracy. These two properties are the most appealing properties of Majorana fermions for topological quantum computation because this state can be manipulated by exchanging them in the language of non-Abelian statistics.

## Jackiw-Rebbi Solutions of Bound States in One Dimension

The simplest model of Majorana fermions in a one-dimensional system is represented by Jackiw-Rebbi solutions [29]. Starting from the Dirac equation for relativistic quantum mechanical wave function which describes an elementary spin ½ particle. The Hamiltonian is

$$H = c\mathbf{p}.\alpha + mc^2\beta. \tag{26}$$

Where $m$ is the rest mass of particle, $\mathbf{p}$ is momentum, and $c$ is the speed of light. $\alpha_i$ and $\beta$ are known as the Dirac matrices and they satisfy anti-commutation relation which means that they obey Clifford algebra and must be expressed in matrix. From this equation, the relativistic energy-momentum relation will be the solution of the equation

$$E^2 = m^2c^4 + p^2c^2. \tag{27}$$

In three dimensions, one has two solutions for positive energy $E_+$ and two solutions for negative energy $E_-$ in the form of

$$E_\pm = \pm\sqrt{m^2c^4 + p^2c^2}. \tag{28}$$

This equation describes the motion of an electron with spin. Two solutions of positive energy for two states of electron with spin-up and spin-down, while two solutions with negative energy solutions for a positron with spin-up and spin-down. The equation demands for the existence of anti-particle i.e., the particle with negative energy or mass. Under the transformation of mass $m$ into $-m$, it is found that this equation remains invariant if β is replaced by $-\beta$ which satisfies all the anti-commutation relations of $\alpha_i$ and $\beta$. It reflects the symmetry between positive and negative energy particles in the Dirac equation. In other words, there is no topological distinction between particles with positive and negative energy masses.

This solution of interface between positive and negative energy masses is introduced in the relation of Dirac equation and topological insulator. Starting with the Hamiltonian,

$$h(x) = -iv\hbar\partial_x\sigma_x + m(x)v^2\sigma_z \tag{29}$$

with

$$m(x) = \begin{cases} -m_1 & \text{if } x < 0 \\ +m_2 & \text{otherwise} \end{cases} \tag{30}$$

($m_1$ and $m_2 > 0$). Effective $v$ is used to replace the speed of light $c$ when the Dirac equation is applied to solid systems. The eigenvalue of the equation is

$$\begin{pmatrix} m(x)v^2 & -iv\hbar\, \partial_x \\ -iv\hbar\, \partial_x & -m(x)v^2 \end{pmatrix} \begin{pmatrix} \varphi_1(x) \\ \varphi_2(x) \end{pmatrix} = E \begin{pmatrix} \varphi_1(x) \\ \varphi_2(x) \end{pmatrix}. \tag{31}$$

For either $x < 0$ or $x > 0$, the equation is a second-order ordinary differential equation. One can solve the equation at either $x < 0$ or $x > 0$ separately and the continuous wave function should appear at $x = 0$. Dirichlet boundary condition is used so that the wave function vanish at $x = \pm\infty$.

At $x = 0$, the continuity condition for the wavefunctions require that

$$\begin{pmatrix} \varphi_1^+ \\ \varphi_2^+ \end{pmatrix} = \begin{pmatrix} \varphi_1^- \\ \varphi_2^- \end{pmatrix} \tag{32}$$

since

$$\varphi_1^+ = -\frac{iv\hbar\lambda_+}{m_2 v^2 - E}\varphi_2^+ \text{ with } \lambda_+ = \pm\sqrt{m_2^2 v^4 - E^2}/v\hbar \tag{33}$$

$$\varphi_1^- = -\frac{iv\hbar\lambda_-}{m_1 v^2 + E}\varphi_2^- \text{ with } \lambda_- = \pm\sqrt{m_1^2 v^4 - E^2}/v\hbar. \tag{34}$$

From this equation, it results in

$$\frac{-\sqrt{m_2^2 v^4 - E^2}}{m_2 v^2 - E} = \frac{\sqrt{m_1^2 v^4 - E^2}}{-m_1 v^2 - E}. \tag{35}$$

Therefore, the solution of zero energy $E = 0$ and its corresponding wave function is

$$\Psi(x) = \sqrt{\frac{v}{\hbar}\frac{m_1 m_2}{m_1 + m_2}} \begin{pmatrix} 1 \\ i \end{pmatrix} e^{-|m(x)vx|/\hbar}. \tag{36}$$

The solution dominantly distributes over the interface or domain wall at x = 0 and decays exponentially away from the original point x = 0. The solution of $m_1 = m_2$ was first obtained by Jackiw and Rebbi [30]. Now, it is a mathematical basis for the existence of topological excitations or solitons in one-dimensional system. If one regards the vacuum as a system with an infinite positive mass, a system of negative mass with an open boundary condition possesses a bound state near the boundary. This result leads to the formation of the edge state and surface states in topological insulators. The solution is quite robust against the mass distribution m(x). If $m(+\infty)$ and $m(-\infty)$ differ by a sign as domain wall, there always exists a zero-energy solution near a domain wall of the mass distribution $m(x)$.

This solution is the origin of Fu and Kane's theoretical proposal [31] of Majorana fermions in surface states of a strong topological insulator when brought into contact with an ordinary s-wave superconductor. It gave rise to proximity-induced superconductivity in topological insulator which has the necessary ingredient of strong spin-orbit coupling i.e., the interaction between the electron's spin and its orbital motion around nucleus.

## Majorana Fermions in Spinless p-wave Superconductor

Prior to the Fu and Kane's theoretical proposal, the isolated Majorana fermion is predicted to occur in vortices and edges of effectively spinless superconducting systems with triplet pairing symmetry, the $p$-wave pairing symmetry in one dimension and $p_x \pm ip_y$ pairing symmetry in two dimension. However, the triplet pairing is very sensitive to disorder and has never been observed experimentally. Fortunately, they may exist in all systems with such topological properties, because Majorana fermion is topologically invariant.

The Hamiltonian in 1D with $p$-wave pairing is first introduced by Kitaev [21]. It has eigenstates of spatially isolated Majorana fermion. Starting with a simple Hamiltonian which describes a spinless $p$-wave superconductor,

$$H_{chain} = -\mu \sum_{i=1}^{N} n_i - \sum_{i=1}^{N-1}\left(t c_i^\dagger c_{i+1} + \Delta\, c_i c_{i+1} + h.c.\right) \tag{37}$$

where $h.c.$ means Hermitian conjugate, $\mu$ is the chemical potential, $c_i$ is the electron annihilation operator for site $i$, and $n_i = c_i^\dagger c_i$ is the associated number operator. The superconducting gap $\Delta$ and hopping amplitude $t$ are assumed to be the same for all sites. Choosing the superconducting phase ϕ to be zero such that $\Delta = |\Delta|$. Assuming that the time-reversal symmetry is broken, it suppresses the spin-label and results in spinless electrons. Moreover, the superconducting pairing is non-standard because it couples electrons with the same spin (in $s$-wave pairing, it couples electrons with opposite spins). The electrons in this manner are paired with neighboring sites. The sites cannot be doubly occupied by the spinless electrons due to Pauli exclusion principle.

Rewriting the Hamiltonian in terms of Majorana operator will result in Majorana fermion which is the splitting of a fermion into its real and imaginary parts. Thus,

$$c_i = \frac{1}{2}(\gamma_{i,1} + i\gamma_{i,2}) \tag{38}$$

$$c_i^\dagger = \frac{1}{2}(\gamma_{i,1} - i\gamma_{i,2}) \tag{39}$$

where $\gamma_{i,j}$ are Majorana operators living on site $i$. They are indeed Majorana operators described by

$$\gamma_{i,1} = c_i^\dagger + c_i \tag{40}$$

$$\gamma_{i,2} = i(c_i^\dagger - c_i). \tag{41}$$

Fig. 3 indicates Kitaev's chain, the upper picture shows fermion operators split into Majorana operators and the lower panel shows two unpaired Majorana operators $\gamma_{1,2}$ and $\gamma_{N,1}$ which can be combined to form zero energy with highly non-local property of fermion operator, $\widetilde{c_M}$, in the limit of $\mu = 0, t = \Delta$. When $\mu = 0, t = \Delta$, by inserting both fermion operators into the Hamiltonian, the result is

$$H_{chain} = -it \sum_{i=1}^{N-1} \gamma_{i,2}\gamma_{i+1,1} \tag{42}$$

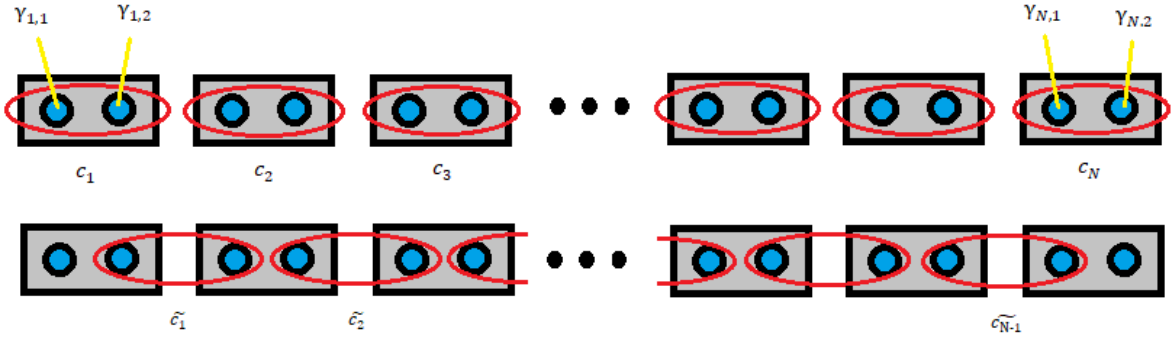

**FIGURE 3.** Animation of Kitaev's 1D $p$-wave superconducting tight binding chain.

This equation is an alternative way of writing diagonalized Hamiltonian. Going back to the fermionic representation where a fermion on site $i$ is split into two Majorana operators living on site $i$, one can construct new fermion operator, $\tilde{c}_\iota$, by combining Majorana operators on neighboring sites

$$\tilde{c}_\iota = (\gamma_{i+1,i} + i\gamma_{i,2})/2. \tag{43}$$

This pairing symmetry is described in lower panel. In terms of this new operators, $-i\gamma_{i,2}\gamma_{i+1,1} = 2\tilde{c}_\iota^\dagger \tilde{c}_\iota = 2\tilde{n}_\iota$ and thus

$$H_{chain} = 2t \sum_{i=1}^{N-1} \tilde{c}_\iota^\dagger \tilde{c}_\iota \tag{44}$$

$\tilde{c}_\iota$ are the annihilation operators corresponding to the eigenstates and energy cost of creating a $\tilde{c}_\iota$ fermion is $2t$. However, the Majorana operators $\gamma_{N,2}$ and $\gamma_{1,1}$ which are localized at the two ends of the wire are completely missing from the equation (42). These two Majorana operators can equivalently be described by a single fermionic state with operator

$$\widetilde{c_M} = (\gamma_{N,2} + i\gamma_{1,1})/2. \tag{45}$$

This is highly non-local state because $\gamma_{N,2}$ and $\gamma_{1,1}$ are localized on opposite ends of the chain. Since this fermion operator is absent from the Hamiltonian, occupying the corresponding state requires zero energy. In contrast to normal superconductor, the ground state is non-degenerate and consists of a superposition of even-particle-number states (Cooper pairs condensate), the Hamiltonian in equation (37) allows for an odd number of quasiparticles existence at zero energy. Again, this is for the very special conditions where $\Delta = t$ and $\mu = 0$. Nevertheless, one can show Majorana end states remain to exist as the chemical potential lies within the gap of $|\mu| < 2t$. The Majorana fermions remain at zero energy as long as the wire is long enough that they do not overlap.

The Hamiltonians for the continuum version of $p$-wave superconductor in 1D and 2D are [32]:

$$H_{1D}^{pw} = \int dx \left[\Psi^\dagger(x)\left(\frac{p_x^2}{2m} - \mu\right)\Psi(x) + \Psi(x)|\triangle|e^{i\theta}p_x \Psi(x) + h.c.\right] \quad (46)$$

$$H_{2D}^{pw} = \int dx \left[\Psi^\dagger(\boldsymbol{r})\left(\frac{\boldsymbol{p}^2}{2m} - \mu\right)\Psi(\boldsymbol{r}) + \Psi(\boldsymbol{r})|\triangle|e^{i\theta}(p_x \pm ip_y\; \Psi(\boldsymbol{r}) + h.c.\right] \quad (47)$$

where $\Psi^\dagger(r)$ is the real space creation operator, $p$ is the momentum, $m$ is the effective electron mass, and $\phi$ is the re-introduction of superconducting phase. The Majorana fermions may appear at the edges of the wire and also at transition points between topological and non-topological region.

## ROAD TO TOPOLOGICAL QUANTUM COMPUTER

### Theoretical Proposals

The idea of employing anyons for topological quantum computer is originated from Kitaev [20]. Before coming into the idea of realizing anyons for topological quantum computer, the two-dimensional systems which support such anyonic scheme must be identified. The topological difference between two and three dimensions was first realized by Leinaas and Myrheim in 1977 [33]. Then in 1982, such two-dimensional scheme was the first system experimentally realized by Tsui *et al.* [34] in fractional quantum Hall liquids. The system was resulted from the condensation of the two-dimensional electron gas in a $GaAs - Ga_xAl_{1-x}As$ heterostructure into a new type of collective ground state. The most striking characteristic is that in some of these phases the particles appear to have fractionalized charge in units of the electron charges. A year later, in 1983, the charge fractionalization was explained by Robert Laughlin [35] in the filling fraction $\nu = 1/3$ that emerge as quasiparticles in highly correlated system.

From Laughlin's theory of ground state and low-lying excitations of a fractional quantum Hall state at $\nu = 1/m$ with $m$ is odd, the hallmark of fractional quantum Hall state is that they support excitations with fractional charge and exotic braiding statistics. In the filling fraction of $\nu = 1/3$ plateau, the low-lying excitations about the ground state are quasiparticles with charge $e/3$. If an electron is added to the system, it will break up into three quasiparticles with charge $e/3$ each. This is possible because the system is composed of many electrons and they are all interacting with each other. Low temperatures and strong magnetic field will enhance the effects of electron-electron interactions. Consequently, when electron is added into $\nu = 1/3$ state, the other electrons rearrange themselves such that no excess charge $e$ is found at the location of added electron. Instead, it is energetically favorable to have three extra charges $e/3$. The size of the charge is controlled by the energy gap $\triangle$.

Since then, such system is known as fractional quantum Hall effect (FQHE) which is characterized by a plateau in the Hall resistance. The robust quantization of this Hall resistance in a quantum Hall state is a direct manifestation of the topological nature of the system's ground state. Thoroughly, in high mobility GaAs heterostructure and quantum wells, electrons can be confined to move in a two-dimensional plane. In a perpendicular magnetic field and low temperature, the electrons in 2D electron gas will organize themselves in a topologically invariant state. The most salient manifestation of topological invariance is the quantization of the transverse or Hall resistance $R = h/\nu e^2$, where $\nu$ is a rational number so-called filling factor, $h$ is Planck constant, and $e$ is electron charge. The quantized $R$ is in fact a topological invariant which is independent of the shape or size of the sample, that is why the quantization is exact and become the characteristics for FQHE.

The anyonic nature of those quasiparticles in FQHE was first realized by Halperin [36]. He mentioned that the appearance of fractional statistics is strongly reminiscent to the description of charged particles tied to magnetic flux tubes in two dimensions as introduced by Wilczek [24]. The relation between non-Abelian anyons and FQHE was then explained theoretically by Moore and Read in 1990 [37]. The fractional charge of quasiparticles in the FQHE implies that the quasiparticles obey fractional statistics, that is adiabatic interchange of two identical quasiparticles produces a phase unequal to $\pm 1$ i.e., anyons. The wave functions transform under interchange of quasiparticles as one-dimensional, i.e., the Abelian representation of the braid group. Mathematically, there exist higher dimensional representation of the braid group. In such representation, the wave function of a set of excitations becomes a vector and each exchange of these particles gives a matrix, i.e., non-Abelian action on this vector. Moore and Read [37] argued that fractional quantum Hall systems are the best candidates of such exotic non-Abelian behavior, since fractional statistics is already believed to occur there.

The first quantum Hall state which is indicated to host non-Abelian anyons is the $\nu = 5/2$ state [37]. These anyons are similar to the Ising model, thus they may support universal quantum computer with some supplementary modifications. Regardless from this fact, many compelling theoretical proposals exist for realizing universal quantum computation in such state [38-40]. The truly universal quantum computation scheme is predicted to be hosted by the filling fraction $\nu = 12/5$ [41] anyons which is equivalent to Fibonacci model that support universal quantum computation [25,42]. However, there are no experimental evidence until now and there is a possibility that such system does not support universal quantum computer [25, 42].

The braiding for topological quantum computing scenario can be thought of the way strands of wire or hair are interlaced in a zigzag manner. A topologically distinct braid cannot be transformed into each other without crossing the world lines. It corresponds to unitary matrices that can be used as building blocks for quantum computation. In quantum Hall system, quasiparticles in quantum Hall edge channels can move around localized quasiparticles in the bulk to demonstrate non-Abelian statistics via electrical conductance. In comparison, in the superconducting system, which is widely developed nowadays, the non-Abelian anyons are demonstrated as mid-gap states (zero modes) bound to a defect such as vortex or endpoint of a nanowire. Zero mode means that the particles exchange occurs in degenerate subspace without changing the energy of the system. Such system usually combines spin-orbit coupling, superconductivity, and Zeeman coupling to the electron spin [43]. Because they bound to a defect, they are typically immobile. Consequently, most proposals focus on unitary braiding operations without physically moving the zero-modes in real space [44], rather by using parameter space. Although, there is real-space braiding alternative [45]. The braiding of Majorana zero modes are only ideas so far, the experiment is still underway. In the other hand, the braiding of $\nu = 1/3$ fractional quantum Hall state is claimed to be achieved recently [46].

In recap, a non-Abelian quantum Hall state is a promising route for topological quantum computation and fault-tolerant quantum computation. However, other physical system also holds promise of such scenario as mentioned earlier. This system is envisioned by Read and Green in 1999 [47], the Majorana fermions that have exotic statistics of non-Abelian anyons. They claimed that for the spinless $p$-wave case, the weak pairing phase has a pair wave function that is asymptotically the same as in Moore-Read state (Pfaffian) quantum Hall state. They argued that its other properties such as edge states, quasihole, and toroidal ground state are similar and thus indicating non-Abelian statistics. The strong pairing phase is Abelian states and the transition between the two phases involves a bulk Majorana fermion, where the mass changes its sign at the transition.

In superconducting systems, the Majorana quasiparticles experiment began since 2001 when Kitaev proposed his model on $p$-wave superconductor [21]. In 2008, Liang Fu and Charles Kane [31] made their proposal using 3D topological insulator platform. Four years later, in 2012, Mourik *et al.* [48] claimed that they had 'signatures' of Majorana quasiparticles in semiconductor nanowire. Another proposal was put forward by Shoucheng Zhang's group [49] in 2010. They utilized chiral Majorana and found its signature in 2018 [50, 60].

Majorana fermion in condensed matter systems, Majorana quasiparticle, must satisfy the Dirac equation and its excitation must be its own anti-particle as described in previous section. Both of those conditions are naturally met in topological superconductors. The former condition is satisfied due to its topological nature. The concept of bulk-boundary correspondence invokes topological superconductors to support gapless excitations on the boundaries which is described by the Dirac equation. The later condition is met because the electron and hole excitations are superimposed in the superconducting state which make them indistinguishable. This condition turns superconductor to have particle-hole symmetry with which the topological gapless boundary excitations become Majorana quasiparticles. When topological superconductor is interfaced with a topologically trivial state, e.g. vacuum, the mismatch of topology will arise. This situation cannot be resolved without having a singularity at the boundary. Thus, the singularity is physically realized as gapless boundary states.

The Kitaev model [21] employs a piece of 'quantum wire' on the surface of three-dimensional superconductor. This is a convenient model to describe a simple but rather unrealistic model which exhibits unpaired Majorana fermions. There are two types of phases, these two conditions represent two phases which exist in the model with the similar bulk properties. However, they are different in terms of boundary properties. One of the phases has unpaired Majorana fermions at the ends of the chain. He described the physical realization using $4\pi$ Josephson junction and use it as quantum gates.

In 2008, the proximity effect of s-wave superconductor and surface state of strong topological insulator was studied by Fu and Kane [31]. It resembles a spinless $p_x + ip_y$ superconductor without breaking time-reversal symmetry. This two-dimensional state supports the existence of Majorana bound states at the vortices. The presence of these $2N$ vortices leads to $2^N$-fold ground state degeneracy. Adiabatically arranged these vortices (i.e., braiding) perform non-trivial operations in degenerate ground space. The linear junctions between superconductors mediated by the topological insulator form non-chiral one-dimensional wire for Majorana quasiparticles. The circuits formed by these junctions is claimed to provide a method for creating, manipulating, and fusing Majorana bound states. This proposal requires many fronts of progress for experimental implementation consisting of strong topological insulator with a robust gap and its interface with an appropriate superconductor.

## Experimental Evidence

A pathbreaking work is conducted by Leo Kouwenhoven's group in 2012 [48]. They claimed that the Majorana quasiparticles signature appears in hybrid superconductor-semiconductor nanowire. This work is pioneered from the earlier work by Sau *et al.* [51] which outlined the necessary ingredients for nanowire devices that would accommodate a pair of Majorana. The first ingredient is one-dimensional semiconducting nanowire with strong spin-orbit interaction. The next ingredient is connecting the nanowire to an ordinary s-wave superconductor. The key of the quantum topological order is the coexistence of spin-orbit coupling with proximity-induced $s$-wave superconductivity and an externally induced Zeeman coupling of the spins. For the Zeeman coupling below a critical value, the system is non-topological (proximity-induced) $s$-wave superconductor. In the other hand, for Zeeman coupling above the critical value, the lowest energy excited state inside a vortex is a zero-energy Majorana fermion state. Therefore, the system has entered into a non-Abelian $s$-wave superconducting state via a topological quantum phase transition (TQPT) tuned by the Zeeman coupling.

Majorana fermions can be detected by various measurement including half-integer conductance quantization, non-local tunneling, 4π periodic Josephson effect, and thermal metal-insulator transition [52]. The most well-known method for its detection is tunneling spectroscopy i.e., half-integer conductance quantization. Sau *et al.* [51] proposed the scanning tunneling experiment from the ends of semiconducting nanowire analytically and numerically which previously have been shown that the Majorana modes at the ends of a one-dimensional $p$-wave superconductor lead to distinct signatures in the scanning tunneling microscopy (STM) spectrum. They found that for Zeeman coupling satisfying ($V_Z > \sqrt{\Delta^2 + \mu^2}$) has a zero-bias conductance peak. This zero-bias peak disappears as the Zeeman splitting is reduced to satisfy ($V_Z < \sqrt{\Delta^2 + \mu^2}$). In the theoretical works proposed by Law *et al.* [53], resonant tunneling into the mid-gap state produces a conductance of $2e^2/h$, whereas without this state the conductance vanishes. This proposal is the origin of the pathbreaking work by Kouwenhoven's group.

However, the proposal of this nanowire system [48] is argued by several other researchers in the same year [54] because the magnitude of zero-bias conductance peak (ZBCP) is $\sim 0.1 e^2/h$ which is an order of magnitude lower than the predicted ideal quantized value ($2e^2/h$) [53]. The key experimental observation is the development of a robust sub-gap ZBCP. Moreover, the Majorana-induced ZBCP should appear only beyond magnetic-field-driven topological quantum phase transition (TQPT) characterized by superconducting gap closing. However, there is no clear signature of this gap closing measured by tunneling current. The papers released after this experiment showed that ZBCP can arise in the absence of Majorana bound states. According to theoretical predictions by Lutchyn *et al.* [55], the observation of a ZBCP at finite magnetic field is only a necessary condition for the existence of Majorana quasiparticles. The sufficient condition is to do interference measurement such as fractional Josephson effect which manifest a 4π periodicity in an ac Josephson measurement. This measurement is unfortunately challenging which makes it unlikely to be successful. Thus, Das Sarma *et al.* [54] proposed a method for validating the existence of Majorana bound states with Majorana splitting oscillation. The work established the oscillations depend sensitively on details of the experiment (constant chemical potential and constant density).

The hybrid system involving semiconductor-superconductor has gained particular attention due to ease of realization and a high degree of experimental control. Typically, the experimental signatures of these experiments are ZBCP in tunneling spectra appearing at finite magnetic field. In a confined normal conductor-superconductor system, Andreev reflection gives rise to discrete electron-hole states below the superconducting gap known as Andreev bound states (ABS). The connection between superconducting proximity effect and ABS in semiconductor-superconductor hybrid system makes zero-energy Majorana bound states (MBS) to be understood as a robust merging of ABS at zero energy.

The existence of Majorana zero mode is demonstrated in various schemes of superconductors coupled with topological matter involving ferromagnetic atomic chains on a superconductor [56], HgTe topological insulator Josephson junctions [57], planar semiconductor heterostructures [58], Fe-based superconductor [59], just to name a few. The detection of its existence is exhibited by zero-bias conductance anomalies modulated by external electrical or magnetic fields. Yet, the detection is rather complicated because of the contributions from other effects, such as Kondo correlations, Andreev bound states, weak anti-localization, and reflection-less tunneling [60]. Conversely, a direct transport signatures of Majorana fermion modes is achieved theoretically and is experimentally observed in quantum anomalous Hall insulator (QAHI) without strong external magnetic field ($\sim 0.1\,T$) and thus preserving superconductivity. By modulating the external field, topological transitions may lead to the establishment of single chiral Majorana edge modes (CMEM) [60].

He *et al.* [60] proposed another platform using hybrid quantum anomalous Hall insulator thin film coupled with a superconductor. Half-integer quantized plateaus $(0.5e^2/h)$ are achieved at the locations of magnetization reversals. The transport measurements revealed that the signature of one-dimensional chiral Majorana fermion modes is reproducible over many magnetic field sweeps at different temperatures. Still, this experimental suggestion is argued recently by Kayyalha *et al.* [61]. They claimed that the half-quantized two-terminal conductance plateau in a millimeter-size QAH-Niobium hybrid device can be realized in similar devices, especially in disordered samples, with a well-controlled and transparent interface as the result of non-Majorana mechanism. The two-terminal conductance is always half-quantized in the strongly coupled superconductor layer and QAH sample with well-aligned magnetization throughout the magnetic field range. The data is obtained from various QAH samples with different geometries demonstrates the robustness, reproducibility, and generality of the presented phenomena. Therefore, the observation of $\sim 0.5e^2/h$ conductance plateau alone is not a sufficient evidence for the existence of CMEM.

The interface between the topological insulator and the superconducting layer is crucial [61]. The authors measured devices where the transport signatures and interface characteristics could be extracted at the same time. The transport signatures previously associated with the chiral Majorana states existed whenever there was a transparent interface between the topological insulator and the superconductor. This strong coupling at the interface allows the superconductor to act as a short, modifying the transport properties from the theoretical predictions. This does not rule out the possibility that the old experiments observed chiral Majorana modes, but it certainly shows that more evidence is needed.

Prior to this argument, the topological quantum computation based on chiral Majorana fermions experiment is performed by Lian *et al.* [50]. The experiment showed that the propagation of chiral Majorana fermions lead to the same unitary transformation which is similar to braiding of Majorana zero modes. The proposal uses topologically protected quantum gates at mesoscopic scales which utilizes propagation of 1D chiral Majorana fermion wave packets with purely electrical manipulations instead of MZM. A Corbino ring junction demonstrated single-qubit quantum-gate operations with chiral Majorana fermion and the conductance of the junction naturally provides readout for the qubit state. This conductance oscillation in the Corbino junction offers the validation of quantum coherent chiral Majorana fermions if observed. However, the experiment is still facing difficulties from error-correction of the phase gate and non-demolitional four-Majorana implementation of the controlled-not gate. In contrast, the group claimed that its computation speed can be $10^3$ faster than the currently existing quantum computation schemes. Still, the decoherence occurs in this experiment from two major sources. First, the non-monochromaticity of the incident electron wave packet. Second, the inelastic scattering originated from electron-phonon coupling.

In 2016, Deng *et al.* [62] performed such experiment which differentiates between topological MBS in finite-length wire and ABS in non-topological (trivial) phase. The MBS is referred to as ABS which has a large degree localized at the wire ends and would evolve into topological MBS as the wire becomes longer. They stated that MBS appears in a coupled quantum-dot hybrid-nanowire system. This experiment demonstrated that MBS is coalescing from Andreev bound state (ABS) in a hybrid InAs nanowire with epitaxial Al using quantum dot at the end of the nanowire as tunneling spectrometer. The quantum dot acts effectively as a single barrier in certain regime.

A year later, in 2017, Liu *et al.* [63] disputed Deng's experiment. They found that generically ABS may coalesce together forming near-zero-energy midgap states as the Zeeman splitting and/or chemical potential are increased. Although, this condition is mostly satisfied in topological regime below the topological quantum phase transition. There is a situation where the ABS may indeed come together to form zero-energy topological MBS. These two conditions are difficult to distinguish since they produce the similar tunneling transport signatures in tunneling conductance spectroscopy. Liu *et al.* [63] discovered that the conductance associated with the coalesced zero-energy trivial ABS is non-universal and could easily be $2e^2/h$ which is mimicking the quantized topological MBS ZBCP value, even in clean disorder-free system. They established that both cases mentioned before can be thought of as overlapping Majorana zero modes and generic zero-sticking property of ABS arises from the combination of spin-orbit coupling, spin splitting, and superconductivity. Finally, they suggested that more decisive transport measurement must demonstrate the non-local nature of Majorana modes (e.g. observing ZBCP from both ends of the wire, measuring non-local correlations) and their robustness against variables (e.g. barrier height, Zeeman splitting, chemical potential, and other variables) in order to validate the existence of topological Majorana zero modes. Hence, the existence of ZBCP value of $2e^2/h$ is a necessity for Majorana zero modes, but not a sufficiency.

The suggestion from Liu *et al.* [63] to observe ZBCP from both ends motivated Lai *et al.* [64] in 2019 to establish an experiment similar to Deng *et al.* [62] with tunneling spectroscopy carried out from both ends of the wire in same sample (quantum dot only at one end) which may distinguish between ABS and MBS arises from ZBCP through simple examination of correlations between two set of tunneling data. MBS will give rise to correlated ZBCP at both ends. In contrary, if ZBCP only exists in the tunneling from one end (uncorrelated) then it is likely to be arising from ABS. They also consider the effect of embedded quantum dots on the cross conductance measured in the same setup. Such measurements lead to differentiate between ABS and MBS by detecting the topological quantum phase transition (TQPT).

This proposal is then refined by Zhang *et al.* [65] via several measurement techniques consisting of peak-to-dip transition in quantized Majorana conductance, non-local Majorana gate effect, correlation and three-terminal Majorana device, Majorana T-shape device for local density of states (LDOS), Majorana-Fu teleportation, and topological Kondo effect. To reveal the true non-local property of MZM, a measurement can be conducted in a three-terminal device with N-S-N setup (normal conductor-superconductor-normal conductor). Both nanowire's ends can detect two LDOS (local density of states) simultaneously by measuring the $dI_1/dV$ and $dI_2/dV$. MZM always appears in pairs which guarantees two ZBCP and their splitting (Majorana oscillation) should be correlated in all parameter space (gates and magnetic field). The most important requirement is that the superconducting part of the wire needs to be long enough (much longer than the spatial distribution of a trivial Andreev bound state). Otherwise $dI_1/dV$ and $dI_2/dV$ may end up detecting trivial state, mimicking a correlation signature of MBS. Fine-tuning the two tunnel barriers may also lead to ZBCP induced from ABS. Therefore, the robustness of ZBCP correlations must be tested in various magnetic field and voltages on all different gates. If one of the wire ends has a quantum dot or smooth potential inhomogeneity, both ZBCPs may not appear because of localized ABS. Thus, an idea of combining crossed Andreev measurement in a long nanowire device may allow one to correlate the appearance of the Majorana ZBCP with a gap closing, since the localized trivial ABS due to the potential inhomogeneity can disturb gap closing point in the local conductance but not in the non-local conductance. This non-local conductance reveals induced gap information of the entire proximitized nanowire if the wire is longer than the superconducting coherence length.

## Challenge in Finding Majorana Fermion for Topological Quantum Computer

The challenge in finding Majorana fermions generally lies in their transport measurement, especially the non-locality measurement as mentioned in the previous sub-section. In line with non-locality measurement, the braiding experiment remains as a quest for topological quantum computing. Yet, there are several ideas to realize Majorana braiding experiment. For quantum Hall effect platform, quasiparticles in quantum Hall edge channels can move around localized quasiparticles in the bulk via the electrical conductance to demonstrate non-Abelian statistics. In superconducting platform, the Majorana fermions that propagate along the edge of a topological superconductor have conventional fermionic exchange statistics, while the mid-gap states (zero-modes) bound to a defect (vortex or end-point of nanowire) are non-Abelian exchange statistics and typically immobile. Thus, most braiding proposals which demonstrate non-Abelian statistics are unitary braiding operation in parameter space (without physically moving the zero-modes in real space). Although, there might be a real-space braiding alternative.

Those proposals are being conducted in the lab. The major quests now are non-locality measurement and non-Abelian braiding experiment. If one can resolve these problems, the promising nature of Majorana fermion for hardware-level resilience against errors to truly overcome scalability problem of quantum computer and achieve fault-tolerant quantum computing may indeed revolutionize quantum computation.

## SUMMARY

The 'signature' of Majorana fermions existence for topological quantum computing applications emerges in various superconducting system platforms. Despite of the clues, the real Majorana fermions evidence that can be implemented as topological qubit in quantum computer is still lacking. However, the experimental development has expanded rigorously since its first idea in almost two decades ago. Many researchers devote their efforts to replicate and improve each other's works. Two major quests remain unanswered: the non-locality measurement and braiding of Majorana fermion. Nevertheless, the experiments are incessantly performed in labs and an optimistic assumption is willing to bet for topological qubit by a decade from now.

# WHAT'S NEXT

Emphasizing on nanowire platform for realizing Majorana-based topological quantum computers which is a main interest nowadays, Indium-antimonide (InSb) and Indium-arsenide (InAs) nanowires are promising candidates. InSb, in particular, is interesting for its high electron mobility, strong spin–orbit coupling and large Landé g-factor. To braid these Majorana states, scalable nanowire networks with a high degree of interconnectivity are required. The recent publication of Leo Kouwenhoven and his colleagues [66] studied the growth dynamics of in-plane InSb nanowires on $InP(111)B$ substrates. Although there exists a large mismatch between the wires and the substrate, single crystalline transport channels which are free from extended defects are formed due to immediate strain relaxation at the nanowire–substrate interface. These in-plane InSb-based devices exhibit high-quality quantum transport, with long phase-coherence length, a hard superconducting-gap, 2e-Coulomb blockade peaks, and possible Majorana/Andreev signatures.

The next step is to establish Majorana zero modes in these structures by performing key experiments like correlation and Majorana braiding [66]. Topological superconductors have chiral edge modes and their chiral motion might be employed for a braiding operation. The obstacle is that Majorana fermions which propagate along the edge of a superconductor have conventional fermionic exchange statistics. Thus, one needs vortices to overcome the problem. The theoretical proposal for braiding in such system are put forward by Beenakker *et al.* [45]. They proposed the exchange in real space with indirect methods. They used the chiral motion along the boundary of superconductor to braid a mobile vortex in the edge channel with an immobile vortex in the bulk. This measurement scheme is fully electrical and deterministic with edge vortices that are created by a voltage pulse at a Josephson junction.

In quantum anomalous Hall system which is envisioned by Shoucheng Zhang's group [50], the proposal for braiding experiment of such chiral Majorana fermion via its propagation was put forward. Although other experiment [61] argued that their finding in previous work [60] of chiral Majorana edge modes showed a non-smoking gun evidence. This platform is currently pursued by Zhou *et al.* [67]. They proposed a method to perform a braiding-like operation in chiral Majorana fermions coupled with quantum dots or Majorana zero modes, a resonant exchange of chiral Majorana fermions can occur and leads to a non-Abelian braiding-like operation analogous to the braiding of Majorana zero modes. Furthermore, they proposed electrical transport experiment schemes to observe the braiding-like operation on four chiral Majorana fermions and to demonstrate the non-Abelian character in four-terminal devices of the quantum anomalous Hall insulator/topological superconductor hybrid junctions.

In fractional quantum Hall system, each charge $e/4$ quasiparticle contains a zero mode and the exchange of two quasiparticles is a non-Abelian operation on a topological qubit encoded in the zero modes. The recent experiment of braiding anyons was performed by Nakamura *et al.* [46] at the $\nu = 1/3$ state. It suggested that they succeeded to measure conductance oscillations in a Fabry-Perot interferometer which agrees with the theoretically predicted value. Although the observed state does not belong to the non-Abelian exchange statistics, the braiding experiment of such Abelian anyonic scheme is compelling.

Other than those platforms mentioned, non-Abelian statistics candidate is predicted by Faugno *et al.* [68] at the $\nu = 3/7$ state of FQHE platform made from GaAs. Furthermore, the FQHE was realized in Weyl semimetals [69]. The prediction for ZrTe material which host non-Abelian reciprocal braiding of Weyl points in Weyl semimetals were achieved by Bouhon *et al.* [70]. In contrast, the universal gates for topological quantum computer through a hybrid multilayer of chiral topological superconductor thin films was obtained by Luo *et al.* [71]. The topological phase gates are assembled by braiding and with those gates, they found a set of topological universal gates for composite Majorana-Ising type quantum computation. They claimed that encoding quantum information in such machine is more effiecient and substantial than that with Fibonacci anyons. The computation result is easier to be readout by electric signals and so are the inputted data. There are many platforms to realize non-Abelian braiding for topological quantum computations. Several systems have a long roadmap since decades ago. Nevertheless, all of those platforms indicate that topological quantum computer is possible to be realized in the future, though nowadays we have zero topological qubit.

# ACKNOWLEDGEMENTS

The authors declare no competing interest. EHH acknowledges ATTRACT 7556175 and CORE 11352881.